\documentstyle[preprint,tighten,eqsecnum,aps,floats,psfig,epsfig]{revtex}

\begin{document}
\draft
\title{
Critical equation of state of randomly dilute Ising systems
}
\author{
Pasquale Calabrese,$^1$ 
Martino De Prato,$^2$ 
Andrea Pelissetto,$^3$ 
Ettore Vicari$^4$ }
\address{$^1$ Scuola Normale Superiore and  INFN, Piazza dei Cavalieri 7,
 I-56126 Pisa, Italy.}
\address{$^2$ Dipartimento di Fisica, Universit\`a di Roma Tre, and 
              INFN, I-00146 Roma, Italy} 
\address{$^3$ Dip. Fisica dell'Universit\`a di Roma ``La Sapienza" \\
and INFN, P.le Moro 2, I-00185 Roma, Italy}
\address{$^4$
Dip. Fisica dell'Universit\`a di Pisa
and INFN, 
V. Buonarroti 2, I-56127 Pisa, Italy}
\address{
{\bf e-mail: \rm 
{\tt calabres@df.unipi.it},
{\tt deprato@fis.uniroma3.it},
{\tt Andrea.Pelissetto@roma1.infn.it},
{\tt vicari@df.unipi.it}
}}

\date{\today}

\maketitle

\begin{abstract}
We determine the critical equation of state of 
three-dimensional randomly dilute Ising systems,
i.e. of the random-exchange Ising universality class.
We first consider the small-magnetization expansion of 
the Helmholtz free energy in the high-temperature phase.
Then, we apply a systematic approximation scheme of the
equation of state in the whole critical regime,
that is based on polynomial parametric representations
matching the small-magnetization of the Helmholtz free
energy and satisfying a global stationarity condition.
These results allow us to estimate several
universal amplitude ratios, such as
the ratio $A^+/A^-$ of the specific-heat amplitudes.
Our best estimate $A^+/A^-=1.6(3)$ is
in good agreement with experimental
results on dilute uniaxial antiferromagnets.

\end{abstract}



\section{Introduction.}
\label{intro}

The critical properties of randomly dilute Ising systems
have been much investigated experimentally and theoretically,
see, e.g., Refs.~\cite{Aharony-76,Stinchcombe-83,Belanger-00,PV-r,FHY-01} 
for reviews.  The typical example is the ferromagnetic random Ising model (RIM) 
with Hamiltonian
\begin{equation}
{\cal H} = -J \sum_{<ij>}  \rho_i \,\rho_j \; s_i s_j
- H \sum_i \rho_i s_i,
\label{rimH}
\end{equation}
where $J>0$, the first sum extends over all nearest-neighbor sites,
$s_i=\pm 1$ are Ising spin variables,
and $\rho_i$ are uncorrelated quenched random variables, which are equal to one 
with probability $p$ (the spin concentration) and zero with probability $1-p$
(the impurity concentration). 
Considerable work has been dedicated to the identification of the 
universal critical behavior of the RIM, and in particular to the 
determination of the critical exponents, which have been 
computed with great accuracy in experiments and in theoretical works.
On the other hand, much less is known about the critical equation of state and 
the corresponding universal amplitude ratios.
From the experimental side, this is essentially due to the fact
that typical experimental realizations of the RIM are
uniaxial antiferromagnets such as 
Fe$_x$Zn$_{1-x}$F${}_2$ and  Mn$_x$Zn$_{1-x}$F${}_2$ materials,
which are usually modeled by the Hamiltonian (\ref{rimH}) with $J<0$.
For $H=0$, there is a simple mapping between the ferromagnetic and 
the antiferromagnetic model, so that the critical behavior of the 
antiferromagnets can still be obtained from that of the ferromagnetic 
RIM. The situation is more complex for $H\not = 0$. The RIM (\ref{rimH}) 
with $H\not=0$ corresponds to an antiferromagnet with a 
staggered magnetic field that cannot be realized
experimentally. Conversely, antiferromagnets in a uniform magnetic field
have a different critical behavior and belong to the same 
universality class of the ferromagnetic random-field Ising model 
\cite{rfimcross}. For these reasons the equation of state of the RIM 
is not relevant for dilute uniaxial antiferromagnets.
However, from the equation of state 
one can derive amplitude ratios involving quantities defined in the high- and 
low-temperature phase that can be measured experimentally,
see, e.g., Ref.~\cite{Belanger-00}.
For example, the ratio $A^+/A^-$ of the specific-heat amplitudes in the 
high- and low-temperature phase has been determined:
$A^+/ A^-=1.6(3)$ (Ref.~\cite{BCSJBKJ-83}),
and $A^+/ A^-=1.55(15)$ (Ref.~\cite{Belanger-00}).
On the theoretical side, only a few works,
based on field-theoretical (FT) perturbative expansions, 
have attempted to determine other universal quantities beside the
critical exponents \cite{GSM-77,Newlove-83,Shpot-90,BS-92,Mayer-98},
obtaining rather imprecise results.
For example, these calculation have not even been able to determine reliably
the sign of the ratio $A^+/A^-$.
Actually, $\epsilon$-expansion calculations \cite{Newlove-83,Shpot-90}
favor a negative value, in clear disagreement with experiments.

In this paper we determine the equation 
of state of the RIM in the critical region, that is the relation 
among the external field $H$,
the reduced temperature $t\equiv (T-T_c)/T_c$, and the magnetization 
\begin{equation} 
M\equiv {1\over V} \overline{ \langle \sum_i \rho_i s_i \rangle },
\end{equation}
where the overline indicates the average over the random variables $\rho_i$,
and $\langle \; \; \rangle$ indicates the sample average at fixed disorder.
It can be written in the usual form as
\begin{equation}
H \propto  M^\delta f(x), \qquad x \propto t M^{1/\beta},
\end{equation}
where $x$ and $f(x)$ are normalized so that $x=-1$ corresponds to the 
coexistence curve, hence $f(-1)=0$, and $f(0)=1$.
In order to determine 
the critical equation of state in the whole critical region,
we use an approximation scheme that has already been applied 
with success to the pure Ising model in three 
\cite{GZ-97,CPRV-99,CPRV-02} and in two dimensions \cite{CHPV-01}.
We first consider the expansion of the equation of state 
in terms of the magnetization in the high-temperature phase.  
The first few nontrivial coefficients can  be determined either from 
Monte Carlo simulations of the RIM or from the analysis of FT perturbative 
expansions.  These results are then used to construct approximations
that are valid in the whole critical
region and that allow us to determine several universal amplitude
ratios.  For example, we anticipate our best estimate of the specific-heat
amplitude ratio 
\begin{equation}
A^+/A^-=1.6(3), 
\end{equation}
which compares very well with the experimental determinations 
in dilute uniaxial antiferromagnets \cite{BCSJBKJ-83,Belanger-00}.

The paper is organized as follows.
In Sec.~\ref{CES} we discuss the general properties of the
critical equation of state. 
In Sec.~\ref{smallMbeh} we consider its 
small-magnetization expansion in the high-temperature phase,
reporting estimates of the first few nontrivial terms,
obtained by Monte Carlo simulations and FT methods.
These results are used in
Sec.~\ref{appeq} to construct approximate polynomial parametric representations
that are valid in the whole critical region and that 
allow us to achieve a rather accurate determination 
of the scaling function $f(x)$.
Finally, in Sec.~\ref{univratio} we determine several universal 
amplitude ratios, such as the specific-heat amplitude ratio
$A^+/A^-$.
In  App.~\ref{notations} we report the definitions of 
the thermodynamic quantities that are considered in the paper.
In App.~\ref{replica} we discuss the correspondence
between the RIM correlation functions and the correlation functions
of the corresponding translation-invariant field theory that is obtained
by using  the standard replica trick.
App.~\ref{specheat} reports some details on the six-loop FT
calculation of the universal amplitude ratio $R_\xi^+$.

\section{The critical equation of state}
\label{CES}

The equation of state relates the 
magnetization $M$, the magnetic field $H$, and the reduced temperature 
$t\equiv (T-T_c)/T_c$. 
In the neighborhood of the critical point $t=0$, $H=0$, it can be 
written in the scaling form 
\begin{eqnarray}
&&H = B_c^{-\delta}M^{\delta} f(x), \\
&&x \equiv t (M/B)^{-1/\beta},
\label{eqstfx}
\end{eqnarray}
where $B_c$ and $B$ are the amplitudes of the magnetization on the critical 
isotherm and on the coexistence curve, see App.~\ref{notations}. 
According to these normalizations, the coexistence curve corresponds to $x=-1$,
and the universal function $f(x)$ satisfies $f(-1)=0$ and $f(0)=1$.
The apparently most precise estimates of the critical exponents
have been recently  obtained by Monte Carlo simulations of 
the RIM: Ref.~\cite{CMPV-03} obtains
$\nu=0.683(3)$ and $\eta=0.035(2)$, while Ref.~\cite{BFMMPR-98}
reports $\nu=0.6837(53)$ and $\eta = 0.0374(45)$.
The FT estimates $\nu=0.678(10)$ and $\eta=0.030(3)$, obtained by
analyzing six-loop fixed-dimension series \cite{PV-00}, are in good agreement
\cite{footnote-FT}.
The other exponents $\alpha$, $\gamma$, $\beta$, and $\delta$ can be determined
using scaling and hyperscaling relations.

The equation of state is analytic for $|H|>0$, implying that
$f(x)$ is regular everywhere for $x>-1$. 
In particular, $f(x)$ has a regular expansion in powers of $x$
around $x=0$
\begin{equation}
f(x) = 1 + \sum_{n=1}^\infty f_n^0 x^n.
\label{expansionfx-xeq0}
\end{equation}
At the coexistence curve, i.e., for $x\rightarrow -1$,
$f(x)$ is expected to have an essential singularity  \cite{singatcoexcurve}, 
so that it can be asymptotically expanded as
\begin{equation}
f(x) \approx  \sum_{n=1}^\infty f^{\rm coex}_n (1+x)^n. 
\label{fxcc} 
\end{equation}
The free energy of a dilute ferromagnet is 
nonanalytic at $H=0$ for all temperatures below the 
transition temperature of the pure system 
\cite{Griffiths-68}, which is larger than the critical temperature
of the dilute one.
Therefore, at variance with pure systems, the equation of state of the RIM
is also nonanalytic for $t>0$ and $H=0$.
However, as argued in Ref.~\cite{Harris-75}, these singularities are
very weak and all derivatives
remain finite at $H=0$, as it happens at the coexistence curve.
This allows us to write down an asymptotic large-$x$ expansion of the form
\begin{equation}
f(x) = x^\gamma \sum_{n=0}^\infty f_n^\infty x^{-2n\beta}.
\label{largexfx}
\end{equation}
It is useful to rewrite the equation of state in terms of a variable 
proportional to $M t^{-\beta}$, although in this case we must distinguish 
between $t > 0$ and $t < 0$. For $t > 0$ we define
\begin{eqnarray}
&& H = \left({C^+\over C_4^+}\right)^{1/2} t^{\beta \delta} F(z),
\nonumber \\
&& z \equiv \left[- {C_4^+\over (C^+)^3} \right]^{1/2} M t^{-\beta},
\label{defFz}
\end{eqnarray}
while for $t<0$ we set
\begin{eqnarray}
&& H = {B\over C^-} (-t)^{\beta \delta} \Phi(u), 
\nonumber \\ 
&& u \equiv {M\over B} (-t)^{-\beta}.
\label{defPhiu}
\end{eqnarray}
The constants $C^\pm$ and $C^+_4$ are the amplitudes appearing in the 
critical behavior 
of the two- and four-point susceptibilities $\chi$ and $\chi_4$
respectively,
see App.~\ref{notations}.
With the chosen normalizations \cite{PV-r},
\begin{eqnarray}
F(z) &=&
z + {1\over 6} z^3
       + \sum_{j=3} {1\over (2j-1)!} r_{2j}\, z^{2j-1},
\label{Fzexpa} \\
\Phi(u) &=& (u-1) + \sum_{j=3}^\infty {1\over (j-1)!} v_{j}\, (u-1)^{j-1}.
\label{phiuexp}
\end{eqnarray}
The large-$z$ expansion of the scaling function $F(z)$ 
is given by
\begin{equation}
F(z) = z^\delta \sum_{k=0} F^{\infty}_k z^{-k/\beta}.
\label{asyFz}
\end{equation}
The functions $F(z)$ and $\Phi(u)$ are clearly related to $f(x)$.
Indeed,
\begin{eqnarray}
&& z^{-\delta} F(z) = F_0^\infty f(x), \qquad\qquad  z = z_0
 x^{-\beta}, \label{rel-Fz-fx}\\
&& u^{-\delta} \Phi(u) = {C^- B^{\delta-1}\over B_c^\delta} f(x),
     \qquad u = (-x)^{-\beta}.
\label{rel-Phiu-fx}
\end{eqnarray}
where $z_0=(R_4^+)^{1/2}$ is a universal constant, see Sec.~\ref{univratio}.

In order to  determine the critical equation of state, we first consider
its small-magnetization expansion.  Then, we construct parametric 
representations of the critical equation of state based on 
polynomial approximations, which are valid in the whole critical
region. This method have already been applied to the Ising universality class 
in three \cite{GZ-97,CPRV-99,CPRV-02} and two dimensions \cite{CHPV-01}, 
and to the three-dimensional $XY$ and Heisenberg universality classes 
\cite{CPRV-00-2,CHPRV-01-02}.

\section{Small-magnetization expansion  in the high-temperature phase} 
\label{smallMbeh}

\subsection{Small-magnetization expansion  of the Helmholtz free energy}
\label{smallhfe}

In the high-temperature phase the quenched Helmholtz free energy
${\cal A}(t,M)$ admits  an expansion around $M=0$:
\begin{equation}
\xi^3 \left[ {\cal A}(t,M) - {\cal A}(t,0)\right] = 
{1\over 2} \hat{M}^2  + 
\sum_{j=2} {1\over (2j)!} G_{2j} \hat{M}^{2j} , 
\label{fe}
\end{equation}
where $\xi$ is the second-moment correlation length along the axis 
$H=M=0$, $\hat{M} \equiv c M \xi^{\beta/\nu}$, 
$c\equiv (f^+)^{1-\eta/2}/(C^+)^{1/2}$, 
and $C^+$, $f^+$ are the amplitudes of $\chi$
and $\xi$ respectively, see App.~\ref{notations} for notations.
We recall that the equation of state is related to the Helmholtz
free energy by
\begin{equation}
H = {\partial {\cal A}(t,M)\over \partial M},
\end{equation}
and therefore the expansion (\ref{fe}) is strictly related to
the expansion of the function $F(z)$, cf. Eq.~(\ref{defFz}).

In the critical limit the coefficients
$G_{2j}$ of the expansion (\ref{fe}) are universal.
They can be determined from the high-temperature critical limit
of combinations of zero-momentum connected
correlation functions averaged over the random dilution
\begin{equation}
\chi_n = \left. {\partial^{n-1} M\over \partial H^{n-1}}\right|_{H=0} \; .
\end{equation} 
Indeed, 
\begin{eqnarray}
&& G_4 = \lim_{t\rightarrow 0^+}\;\lim_{V\rightarrow \infty}\;
\left( - {\chi_4 \over \xi^d \chi_2^{\,2}} \right),
\label{Gkdef}
\\
&& G_6 = \lim_{t\rightarrow 0^+}\;\lim_{V\rightarrow \infty}\;
\left( - { \chi_6 \over \xi^{2d} \chi_2^{\,3}}
+ { \chi_4^{\,2} \over \xi^{2d} \chi_2^{\,4}}\right) ,
\nonumber \\
&&G_8 = \lim_{t\rightarrow 0^+}\;\lim_{V\rightarrow \infty}\;
\left( - { \chi_8 \over \xi^{3d} \chi_2^{\,4}}
+ 56 { \chi_6 \chi_4 \over \xi^{3d} \chi_2^{\,5}} 
- 280 { \chi_4^{\,3} \over \xi^{3d} \chi_2^{\,6}}\right),
\nonumber
\end{eqnarray}
where $\chi_2 \equiv \chi$ and $V$ is the volume.

For the purpose of determining the small-magnetization
expansion of the RIM equation of state, it 
is convenient to rewrite the Helmholtz free energy of the RIM, 
cf. Eq.~(\ref{fe}), in the equivalent form
\begin{equation}
{\cal A}(t,M) - {\cal A}(t,0)= - { (C^+)^2\over C_4^+} t^{3\nu} A(z),
\end{equation}
where $z$ is defined in Eq.~(\ref{defFz}), 
\begin{equation}
A(z) =  {1\over 2} z^2 + {1\over 4!} z^4 + 
\sum_{j=3} {1\over (2j)!} r_{2j} z^{2j},
\label{AZ}
\end{equation}
and
\begin{equation}
r_{2j} = {G_{2j}\over G_4^{j-1}} \qquad\qquad j\geq 3.
\label{r2j}
\end{equation}
Since $F(z)={dA(z)/dz}$, Eq.~(\ref{defFz}) follows from Eq.~(\ref{AZ}).

Some of these universal constants have been recently estimated by means of a
Monte Carlo simulation \cite{CMPV-03}, obtaining
\begin{equation}
G_4=43.3(2),\qquad r_6=0.90(15).
\label{MCres}
\end{equation}
Correspondingly, $G_6=1.7(3)\times 10^3$.

\subsection{Field-theoretical approach}
\label{FTappr}

One may estimate the universal quantities $G_{2j}$ and $r_{2j}$ by FT methods.
The FT approach is based on 
an effective Landau-Ginzburg-Wilson  Hamiltonian \cite{replica} that can
be obtained by using the replica method, 
\begin{equation}
{\cal H}_{\varphi^4} =  \int d^3 x \left\{ {1\over 2} \sum_{i=1}^{N}
      \left[ (\partial_\mu \varphi_i)^2 +  r \varphi_i^2 \right]  
+{1\over 4!} \sum_{i,j=1}^N \left( u_0 + v_0 \delta_{ij} \right)
+ H\sum_{i=1}^N \varphi_i  
\right\} \; ,
\label{Hphi4rim}
\end{equation}
where $\varphi_i$ is an $N$-component field.
The critical behavior of the RIM is expected to be described
by the Hamiltonian ${\cal H}_{\varphi^4}$ for $u_0<0$ and
in the limit $N\rightarrow 0$.
Conventional FT computational schemes show that the fixed point
corresponding to the pure Ising model is unstable and that 
the renormalization-group
flow moves towards another stable fixed point describing the critical behavior 
of the RIM.

The most precise FT results for the critical exponents
have been obtained in the framework of the perturbative fixed-dimension
expansion in terms of zero-momentum couplings. 
 The corresponding Callan-Symanzik $\beta$-functions and the 
renormalization-group functions
associated with the critical exponents have been computed to
six loops \cite{PV-00,CPV-00}.

The Helmholtz free energy ${\cal A}_{\varphi^4}$
associated with the Hamiltonian (\ref{Hphi4rim})
can be written as \cite{PS-01,PS-02}
\begin{eqnarray}
&& \xi^3 \left[ {\cal A}_{\varphi^4}(M)-{\cal A}_{\varphi^4}(0)\right] =
{1\over 2} \sum_a \hat{M}_a^2  + 
{1\over 4!} \sum_{ab} ( g_{41} + g_{42} \delta_{ab}) \hat{M}_a^2 \hat{M}_b^2 + 
\label{fec}\\&&
+{1\over 6!} \sum_{abc} ( g_{61} + g_{62} \delta_{ab} + g_{63} \delta_{abc})
 \hat{M}_a^2 \hat{M}_b^2 \hat{M}_c^2 +
\nonumber
\\
&& 
+{1\over 8!} \sum_{abcd} \left( g_{81} + g_{82} \delta_{ab} + 
g_{83} \delta_{ab}\delta_{cd} +
g_{84} \delta_{abc} 
+ g_{85} \delta_{abcd} \right)
 \hat{M}_a^2 \hat{M}_b^2 \hat{M}_c^2 \hat{M}_d^2 + ...
\nonumber
\end{eqnarray}
where $\hat{M}_a = c M_a\xi^{\beta/\nu}$, 
$c = \xi^{1 - \eta/2}/\chi^{1/2}|_{H=0}$.
Note that $g_{41}=u^*$ and $g_{42}=v^*$,
where $u^*$ and $v^*$ are the fixed-point values of the
renormalized quartic couplings $u$ and $v$.
Applying the replica method in the presence of a uniform
external field $H$, see App.~\ref{replica} for details, 
one can identify
\begin{eqnarray}
G_4 = g_{42} = v^*, \qquad G_{6} = g_{63}, \qquad G_8 = g_{85},
\end{eqnarray}
etc...
It is worth noting that 
the other coefficients appearing in the
small-magnetization expansion of the Helmholtz free energy (\ref{fec}), i.e.
$g_{41}$, $g_{61}$, $g_{62}$, etc..., can be related to 
dilution averages of products of single-sample $n$-point correlation
functions, as shown in App.~\ref{replica}.

The quartic coupling $G_4$  can be estimated
from the position  of the RIM fixed point in the $u$, $v$ plane,
i.e. from the common zero of their $\beta$-functions.
The analysis of the six-loop perturbative series gives results 
somewhat dependent on the resummation method \cite{PV-00},
see also the five-loop analysis of Ref.~\cite{PS-00}. 
Combining all results together, one finds $36.5 \lesssim G_4 \lesssim 39.5$,
which can be summarized as $G_4 = 38.0(1.5)$. Such a result
is not consistent with the Monte Carlo estimate (\ref{MCres}),
obtained by estimating the limit given in Eq.~(\ref{Gkdef}).
But, as discussed in Ref.~\cite{CMPV-03}, the quantitative consequences 
for the determination of the critical exponents
are very small, since the renormalization-group functions corresponding
to the critical exponents turn out to be little 
sensitive to the correct position of the fixed point~\cite{footnoteFP}
(along the Ising-to-RIM renormalization-group
trajectory, see also Ref.~\cite{CPPV-03}).

The universal constants
$G_6$ and $r_6= G_6/G_4^2$ have been estimated in 
Ref.~\cite{PS-02} by analyzing the corresponding
four-loop series with the Pad\'e-Borel method. 
These analyses provide the estimate $r_6\approx 1.09$.
We have reanalyzed this series, finding that, 
unlike critical exponents, $r_6$ is rather sensitive
to the position of the fixed point. Using the Monte Carlo
estimates of $u^*$ and $v^*$ and the Pad\'e-Borel method,
we obtain $r_6\approx 0.6$.
Thus, a conservative estimate is $r_6=1.1^{+0.1}_{-0.5}$,
which is in agreement with the Monte Carlo result of Ref.~\cite{CMPV-03}
mentioned in Sec.~\ref{smallhfe}, $r_6 = 0.90(15)$.
We have also computed the fixed-dimension perturbative expansion
of $r_8\equiv G_8/G_4^3$ to three loops. 
This requires the computation of
the 8-point one-particle irreducible zero-momentum correlation function. 
At three loops this calculation requires the evaluation of 42 
Feynman diagrams. For this purpose we have used the general algorithm
of Ref.~\cite{PV-00-eqst}. The expansion of $r_8$ is
\begin{eqnarray}
&&r_8 = - {105\over 64 \pi} (8 u + 3 v) + {35\over 9216 \pi^2}
(2480 u^2 + 2286 u v + 585 v^2) \label{r8series}\\
&&\;\; - 0.132874 \,u^3 - 0.199366 \,u^2 v - 0.1080052 \,u v^2 -
0.0211475\, v^3 + ...
\nonumber 
\end{eqnarray}
where the zero-momentum quartic couplings $u$ and $v$ are normalized so that
$u = u_0/m$ and $v=v_0/m$ at tree level (see App.~\ref{specheat} for precise
definitions). We have checked that Eq.~(\ref{r8series}) 
correctly reproduces the known series
for the O($N$) models \cite{BBMN-87} with $N=0,1$ in the appropriate limits.
Unfortunately, the analysis of the expansion (\ref{r8series}) 
provides only a very rough estimate \cite{PBanr8},
$r_8=15(15)$.  We will obtain a much better estimate of $r_8$ in 
Sec.~\ref{reseq}, from the results for the equation of state.

\section{Approximate representations of the critical equation of state}
\label{appeq}

\subsection{Polynomial parametric representations}
\label{ppr}

In order to obtain approximate expressions for the equation of state,
we parametrize the thermodynamic variables in terms of two parameters
$R$ and $\theta$, implementing all expected
scaling and analytic properties. Explicitly, we write 
\cite{parrep} 
\begin{eqnarray}
M &=& m_0 R^\beta \theta ,\nonumber \\
t &=& R(1-\theta^2), \nonumber \\
H &=& h_0 R^{\beta\delta}h(\theta), \label{parrep}
\end{eqnarray}
where $h_0$ and $m_0$ are normalization constants.  
The variable $R$ is nonnegative and measures
the distance from the critical point in the $(t,H)$ plane;
the critical behavior is obtained for $R\to 0$.
The variable $\theta$  parametrizes the displacements along the lines
of constant $R$. 
The line $\theta=0$ corresponds to the high-temperature phase $t>0$ and $H=0$;
the line $\theta=1$ to the critical isotherm $t=0$;
$\theta=\theta_0$, where $\theta_0$ is the smallest positive zero
of $h(\theta)$, to the coexistence curve $T<T_c$ and $H\to 0$.
Of course, one should have $\theta_0 > 1$ and $h(\theta)> 0$ for 
$0< \theta< \theta_0$.
The function $h(\theta)$ must be
analytic in the physical interval $0\le\theta<\theta_0$ in order to satisfy the
requirements of regularity of the equation of state (Griffiths' analyticity).
Note that the mapping (\ref{parrep}) is not invertible when
its Jacobian vanishes \cite{PV-r}, which occurs  for
$\theta^2_l=1/(1-2\beta)$.
Thus, a parametric representation 
is acceptable only if $\theta_0<\theta_l$.
The function $h(\theta)$ must be odd in $\theta$,
to guarantee that the equation of state has an
expansion in odd powers of $|M|$ in the 
high-temperature phase for $|M|\to 0$. 
Moreover, it can be normalized so that 
$h(\theta)=\theta+O(\theta^3)$.

The scaling functions $f(x)$ and $F(z)$ can be expressed
in terms of $\theta$.
The scaling function $f(x)$ is obtained from
\begin{eqnarray}
&& x = {1 - \theta^2\over \theta_0^2 - 1} 
\left( {\theta_0\over \theta}\right)^{1/\beta}, \nonumber \\
&& f(x) = \theta^{-\delta} {h(\theta)\over h(1)},
\label{fxmt}
\end{eqnarray}
while $F(z)$ is obtained by
\begin{eqnarray}
&&z = \rho \theta \left( 1 - \theta^2\right)^{-\beta},
\nonumber \\
&&F(z(\theta)) = \rho \left( 1 - \theta^2 \right)^{-\beta\delta} h(\theta),
\label{Fzrel}
\end{eqnarray}
where $\rho$ can be related to $m_0$, $h_0$, $C^+$, and $C_4^+$ by using 
Eqs. (\ref{defFz}) and (\ref{parrep}).

Eq. (\ref{parrep}) and the normalization
condition $h(\theta)\approx \theta$ for $\theta\to 0$ do not completely
fix the function $h(\theta)$. Indeed, one can rewrite
the relation between $x$ and $\theta$ in the form
\begin{equation}
x^\gamma = h(1) \, f_0^\infty (1 - \theta^2)^\gamma 
    \theta^{1-\delta}.
\end{equation}
Thus, given $f(x)$,
the value of $h(1)$ can be arbitrarily chosen to 
completely fix $h(\theta)$. 
One may fix this arbitrariness by choosing arbitrarily the parameter $\rho$  
in the expression (\ref{Fzrel}). 

We approximate $h(\theta)$ with polynomials, i.e., we set
\begin{equation} 
     h(\theta) = \theta + \sum_{n=1}^k h_{2n+1} \theta^{2n+1}.
\label{parametrich}
\end{equation} 
This approximation scheme turned out to be effective in the case
of pure Ising systems \cite{GZ-97,CPRV-99}.
If we require  the approximate parametric representation 
to give the correct $(k-1)$ universal ratios $r_6$, $r_8$, 
$\ldots$, $r_{2k+2}$, we obtain
\begin{equation}
   h_{2n+1} = \sum_{m=0}^n c_{nm} 6^m (h_3 + \gamma)^m {r_{2m+2}\over (2m+1)!},
\label{hcoeff}
\end{equation}
where
\begin{equation}
   c_{nm} = {1\over (n-m)!} \prod_{k=1}^{n-m} 
   (2\beta m - \gamma + k - 1),
\end{equation}
and we have set $r_2 = r_4 = 1$. 
Moreover, by requiring that $F(z)=z + {1\over 6} z^3 + ...$,
we obtain the relation
\begin{equation}
   \rho^2 = 6 (h_3 + \gamma).
\label{h3intermsofrho}
\end{equation}
In the exact parametric representation, the coefficient $h_3$ 
can be chosen arbitrarily. 
This is no longer true when we use our truncated function $h(\theta)$, 
and the related approximate function  $f_{\rm approx}^{(k)}(x,h_3)$ 
depends on $h_3$. We must thus fix a particular value for this parameter. 
In order to optimize this choice, we employ a
variational procedure \cite{CPRV-99},  requiring the approximate function
$f_{\rm approx}^{(k)}(x,h_3)$ to have the smallest possible dependence 
on $h_3$. This is achieved by setting $h_3 = h_{3,k}$, where $h_{3,k}$ is 
a solution of the global stationarity condition
\begin{equation}
  \left. 
  {\partial f_{\rm approx}^{(k)}(x,h_3) \over 
   \partial h_3}\right|_{h_3 = h_{3,k}} 
  = 0
\label{globalstationarity}
\end{equation}
for all $x$.
The existence of such a value of $h_{3,k}$ for each $k$ is
a nontrivial mathematical result which was proved in Ref.~\cite{CPRV-99}. 
This procedure represents a systematic approximation scheme,
which is only limited by the number of known terms in the small-magnetization
expansion of the Helmholtz free energy.
Note, for $k=1$, the so-called linear model, Eq. (\ref{globalstationarity})
gives
\begin{equation}
    h_3 = {\gamma (1 - 2 \beta)\over \gamma - 2 \beta},
\end{equation}
which was considered as the optimal value of $h_3$ 
for the Ising equation of state \cite{SLH-69}.

\subsection{Results}
\label{reseq}

\begin{table}[tbp]
\footnotesize
\caption{
Expansion coefficients for the 
scaling equation of state obtained 
by the $k=1,2$ approximations of the parametric function
$h(\theta)$.
See text for definitions.
The number marked
with an asterisk is an input, not a prediction.  
}
\label{eqstdet0}
\begin{tabular}{cccc}
\multicolumn{1}{c}{}&
\multicolumn{1}{c}{$k=1$}&
\multicolumn{1}{c}{$k=2$}&
\multicolumn{1}{c}{Ising (Ref.~\cite{CPRV-02})}\\
\tableline
$f^0_1$            & 1.100(2)   & 1.06(2)    & 1.0527(7) \\

$f^0_2$            & 0.083(1)   & 0.06(2)    & 0.0446(4) \\

$f^0_3$            & $-$0.012(2)& $-$0.006(3)& $-$0.0059(2) \\

$f_1^{\rm coex}$   & 0.87(1)    &  0.93(8)   & 0.9357(11)\\

$f^\infty_0$       & 0.550(3)   & 0.48(3)    & 0.6024(15)  \\

$r_6$              & 1.46(6)    &$^*$0.90(15)&  2.056(5) \\

$r_8$              & 1.0(3)     & 1.5(3)     &  2.3(1) \\

$v_3$              & 6.9(2)     & 6.3(2)     & 6.050(13) \\

$v_4$              & 17(1)      & 18.7(5)    & 16.17(10) \\

$F^\infty_0$       & 0.0235(6)  & 0.018(2)   & 0.03382(15) 
\end{tabular}
\end{table}

\begin{figure}[tb]
\hspace{-1cm}
\vspace{0.2cm}
\centerline{\psfig{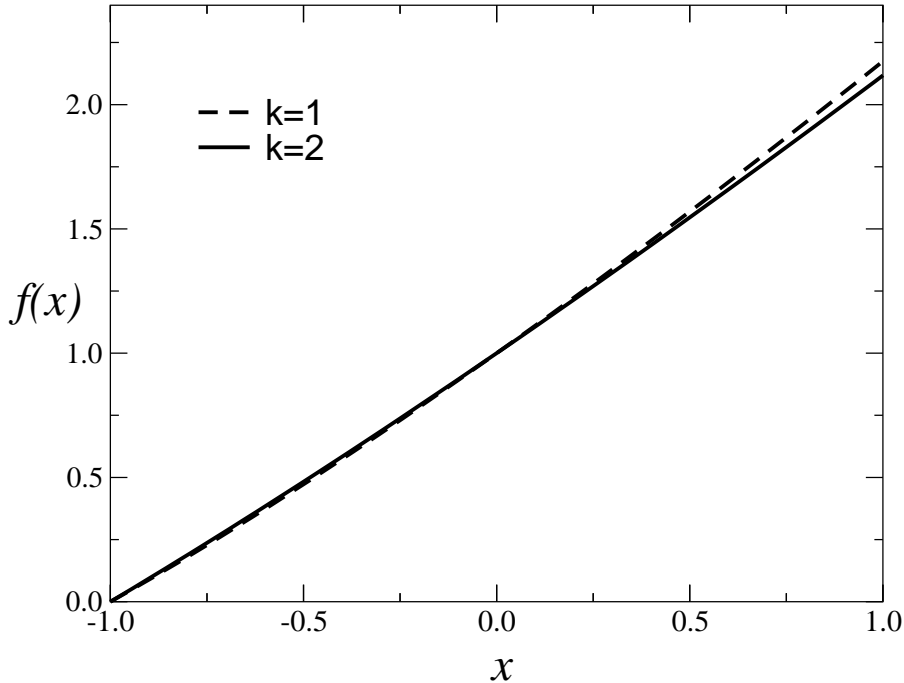}}
\vspace{-0.2cm}
\caption{
The scaling function $f(x)$.}
\label{figfx}
\end{figure}

We apply the method outlined in the preceding section,
using the Monte Carlo estimates \cite{CMPV-03} $\nu=0.683(3)$, $\eta=0.035(2)$,
$r_6=0.90(15)$ as input parameters.
We obtain two different approximations corresponding to $k=1$ and $k=2$.
Using the central values of the input parameters, we have
\begin{equation}
h(\theta)^{(k=1)} = \theta \left( 1 - \theta^2/\theta_0^2 \right),
\qquad \theta_0^2=1.61451,
\end{equation}
that provides the optimal linear model, 
and
\begin{eqnarray}
&&h(\theta)^{(k=2)} = 
\theta \left( 1 - \theta^2/\theta_0^2 \right)
\left( 1 + c_1 \theta^2 \right),\\
&&\theta_0^2=1.32141,\qquad c_1=0.106612.
\end{eqnarray}
The relatively small value of $c_1$ supports the effectiveness
of the approximation scheme.
In Table~\ref{eqstdet0} we report results concerning the behavior 
of the scaling function $f(x)$, $F(z)$, and $\Phi(u)$ for $H=0$ and on the 
critical isotherm, 
cf. Eqs.~(\protect\ref{expansionfx-xeq0}), 
(\protect\ref{fxcc}), 
(\protect\ref{largexfx}), 
(\protect\ref{Fzexpa}), 
(\protect\ref{phiuexp}), (\protect\ref{asyFz}).
The errors reported there are only related to the
uncertainty on the corresponding input parameters.
We consider the $k=2$ results as our best estimates.
Of course, the corresponding errors do not take into account 
the systematic error due to the approximation scheme.
Nevertheless, on the basis of the preceding applications 
\cite{CPRV-99,CPRV-02,CHPV-01}
to the three- and two-dimensional Ising universality class,
we believe that the $k=2$ approximation already provides 
a reliable estimate, and that one may take 
the difference with the $k=1$ result as indicative
estimate of (or bound on) the systematic error.
For comparison, the last column reports the corresponding estimates
for the Ising universality class, taken from Ref.~\cite{CPRV-02}.
In  Fig.~\ref{figfx} we show 
the scaling function $f(x)$, 
as obtained from the $k=1,2$ approximations of $h(\theta)$,
using the central values of the input parameters.
The difference between the two curves is rather small
in the region $-1\le x \lesssim 1$. For larger values of $x$ some differences 
are observed: they are essentially due to the small difference
(approximately 10\%) in the corresponding values of $f_0^\infty$.

\section{Universal amplitude ratios}
\label{univratio}

Universal amplitude ratios characterize the critical behavior
of thermodynamic quantities that do not depend on
the normalizations of the external (magnetic) field, of the order
parameter (magnetization), and of the temperature.  
From the scaling function $f(x)$ one may derive 
many universal amplitude ratios involving quantities
defined at zero momentum (i.e.\ integrated in the volume), such as
the specific heat, the magnetic susceptibility, etc....
For example, one can obtain  estimates for 
the specific-heat amplitude ratio $A^+/A^-$, the 
sueceptibility amplitude ratio $C^+/C^-$, etc....
Then, using the results for $G_4$ and the relation
\begin{equation}
G_4\equiv -{C_4^+\over (C^+)^2 (f^+)^3},
\end{equation}
where $C_4^+$, $C^+$, and $f^+$ are respectively the amplitudes
of the zero-momentum connected four-point correlation,
the susceptibility, and the second-moment correlation length
(see App.~\ref{notations} for notations),
we can also determine universal ratios involving
the correlation-length amplitude $f^+$.

In Table \ref{ratios} we report estimates of several amplitude ratios,
as derived by using the approximate polynomial representations of the 
equation of state for $k=1,2$.
Again, the errors reported there are only related to the
uncertainty on the corresponding input parameters.
Note that, in the most important case of 
$A^+/A^-$, the $k=2$ estimate includes the $k=1$
result within its error.
As discussed in Sec.~\ref{reseq},
we consider the $k=2$ result as our best estimate for each quantity.
As an indicative estimate of the total uncertainty,
we take the maximum between the error induced
by the input parameters and the difference with the $k=1$ result.
This would lead to the final estimates
$A^+/A^-=1.6(3)$, $C^+/C^-=5.5(1.0)$,
$R_c^+=0.08(3)$, 
$R_4^+=12(2)$,  $R_\chi=2.1(3)$ and 
$G_4^{1/3} R_\xi^+=0.99(1)$.

Universal amplitude ratios 
involving the correlation-length amplitude $f^+$, such as 
\begin{equation}
R_\xi^+ \equiv (\alpha A^+)^{1/3} f^+,\qquad 
Q_c \equiv B^2(f^+)^3/C^+,
\end{equation}
can be obtained by using the Monte Carlo estimate of 
Ref.~\cite{CMPV-03}, $G_4=43.3(2)$:
\begin{eqnarray}
&&R_\xi^+ = \left( {R_4^+ R_c^+\over G_4} \right)^{1/3} = 0.282(3), 
\label{rxi}\\
&&Q_c={R_4^+\over G_4} = 0.28(5).
\end{eqnarray}

\begin{table}[tbp]
\footnotesize
\caption{
Universal amplitude ratios obtained from the $k=1,2$ approximations 
of the parametric function $h(\theta)$. 
Amplitude definitions are reported in App.~\protect\ref{notations}.
For comparison, the last column reports the corresponding estimates
for the Ising universality class, taken from Ref.~\protect\cite{CPRV-02}.
}
\label{ratios}
\begin{tabular}{lccc}
\multicolumn{1}{c}{}&
\multicolumn{1}{c}{$k=1$}&
\multicolumn{1}{c}{$k=2$}&
\multicolumn{1}{c}{Ising}\\
\tableline
$A^+/A^-$ 
& 1.33(8) & 1.58(26) &  0.532(3) \\

$C^+/C^-$ 
& 4.5(2) & 5.5(5) & 4.76(2) \\  

$R_c^+\equiv \alpha A^+C^+/B^2$ 
& 0.1008(5) & 0.079(6) & 0.0567(3)  \\

$R_4^+\equiv - C_4^+B^2/(C^+)^3 $ & 
9.9(2) & 12.2(8) & 7.81(2) \\

$R_\chi\equiv C^+ B^{\delta-1}/(B_c)^\delta$ 
& 1.82(1) & 2.1(1) & 1.660(4)\\

$G_4^{1/3} R_\xi^+ \equiv (-\alpha A^+ C_4^+/(C^+)^2)^{1/3}$ 
& 0.999(5) & 0.989(5) & 0.443(2) 
\end{tabular}
\end{table}

The result for $R_\xi^+$ is in substantial agreement with the 
rather precise Monte Carlo  estimate \cite{CMPV-03} $R_\xi^+=0.2885(15)$,
providing support to our approximation of the equation of state.
We have also obtained another independent estimate of the 
ratio $R_\xi^+$ by using the FT fixed-dimension approach in terms
of zero-momentum renormalized coulings.
As discussed in detail in App.~\ref{specheat} we have extended the 
five-loop calculation of Ref.~\cite{Mayer-98} to six loops.
The analysis of the perturbative expansions gives the estimate  
$R_\xi^+=0.290(10)$, which is in good agreement with the 
results from the equation of state and from Monte Carlo simulations.

The comparison with experiments on uniaxial 
antiferromagnets, see, e.g., Ref.~\cite{Belanger-00}
for a review, is essentially restricted to the specific-heat ratio $A^+/A^-$.
Our best estimate  $A^+/A^-=1.6(3)$ is in good agreement
with the experimental result 
$A^+/A^-=1.6(3)$ reported in Ref.~\cite{BCSJBKJ-83},
and $A^+/A^-=1.55(15)$ reported in Ref.~\cite{Belanger-00}.
Earlier FT studies in the framework of $\epsilon$
and fixed-dimension expansions \cite{Newlove-83,Shpot-90,BS-92} 
provided rather imprecise results. 
Actually, they favored a negative value,
in substantial disagreement with experiments.

Other experimental results concern the staggered magnetic susceptibility
and the correlation length, see, e.g., Refs.~\cite{BKJ-86,Belanger-00},
which are determined from neutron-scattering experiments.
As noticed in Ref.~\cite{PA-85},
the two-point correlation function measured in these experiments
does not coincide with
\begin{equation}
G(x) = \overline{\langle \rho_0 s_0 \,\rho_x s_x \rangle_c }=
\overline{\langle \rho_0 s_0 \,\rho_x s_x \rangle }- {1\over V}
\overline{ \langle \sum_i \rho_i s_i \rangle^2 }, 
\end{equation}
whose zero-momentum component is given by the susceptibility 
$\chi=\partial M/\partial H$, but rather with
\begin{equation}
\widehat{G}(x) = \overline{\langle \rho_0 s_0 \,\rho_x s_x \rangle}
- {1\over V} \overline{\langle \sum_i \rho_i s_i \rangle}^2.
\end{equation}
This difference  affects essentially the low-temperature critical behavior.
Indeed, setting 
$\hat{\chi}= \hat{C}^{\,\pm} |t|^{-\gamma}$
for the critical behavior of the corresponding
susceptibility in the high- and low-temperature phase,
$\hat{C}^+=C^+$, but $\hat{C}^-\ne C^-$, and therefore
$C^+/C^-\ne \hat{C}^+/\hat{C}^-$.
A leading-order calculation \cite{PA-85}
within the $\sqrt{\epsilon}$ expansion gives
\begin{equation}
{C^+\over C^-} =
{\hat{C}^+\over \hat{C}^-}
\left[ {7\over 4} + O(\sqrt{\epsilon})\right].
\label{corrf}
\end{equation}
Trusting the leading-order result (\ref{corrf}) and using 
$C^+/C^- = 5.5(1.0)$, see Table~\ref{ratios}, 
we obtain approximately ${\hat{C}^+/ \hat{C}^-} \approx3$, 
which is consistent with the experimental result
${\hat{C}^+/\hat{C}^-}=2.8(2)$ of 
Ref.~\cite{BKJ-86}.

\appendix

\section{Notations}
\label{notations}

In this appendix we define the thermodynamic quantities 
considered in this paper and their critical behavior.
We consider: the specific heat 
\begin{equation}
C_H \equiv {1\over V} 
  \left( \overline{ \langle E^2 \rangle - \langle E \rangle^2}\right),
\label{chdef}
\end{equation}
($E\equiv-J \sum_{<ij>}  \rho_i \,\rho_j \; s_i s_j$),
whose critical behavior is
\begin{equation}
C_H = D  + A^{\pm} |t|^{-\alpha}, 
\label{sphamp}
\end{equation}
where $D$ is a background constant that is the leading contribution for 
$t\to 0$ since $\alpha < 0$;
the spontaneous magnetization near the coexistence curve
\begin{equation}
M \equiv  {1\over V} \overline{ \langle \sum_i \rho_i s_i \rangle }=
B |t|^{-\beta} ;
\label{magamp}
\end{equation}
the magnetic susceptibility $\chi$ and 
the second-moment correlation length $\xi$,
\begin{equation}
\chi = C^{\pm} |t|^{-\gamma}, \qquad \xi = f^{\pm} |t|^{-\nu},
\label{chixiamp} 
\end{equation}
defined from the connected two-point  function
averaged over random dilution
\begin{equation}
G(x) = \overline{\langle \rho_0 s_0 \,\rho_x s_x \rangle_c}\; ;
\end{equation}
the $n$-point susceptibilities $\chi_n$, defined as
the zero-momentum connected $n$-point correlations averaged over 
random dilution, whose asymptotic critical behavior
is written as 
\begin{equation}
\chi_n = C_n^\pm |t|^{-\gamma-(n-2)\beta\delta}.
\end{equation}
We also  consider 
amplitudes defined in terms of the critical behavior along the critical 
isotherm $t = 0$, such as
\begin{equation}
M = B_c H^{1/\delta} , \qquad 
\chi = {B_c\over \delta} |H|^{-{\gamma/\beta\delta}}. \label{def-Bcconst} 
\end{equation}

\section{Correspondence between RIM and FT correlations}
\label{replica}

In this appendix we determine the relationships between
the RIM correlation functions of the spins $s_i$ and those of the 
field $\varphi(x)$ that can be computed in the 
FT approach discussed in Sec.~\ref{FTappr}.
In the RIM, beside the $n$-point susceptibilities $\chi_n$,
defined as the zero-momentum $n$-point connected correlation functions
averaged over random dilution, one may also consider 
dilution averages of products of sample averages.

Setting
\begin{equation}
\sigma_{k} \equiv \left\langle \; ( \sum_i \rho_i s_i\; )^k \right\rangle_c, 
\label{muk}
\end{equation}
we define the dilution-averaged correlations 
\begin{equation}
\rho_{k_1k_2...k_n} \equiv 
  \overline{ \sigma_{k_1} \sigma_{k_2} ...\sigma_{k_n}},
\label{rhodef}
\end{equation}
and the generalized susceptibilities
\begin{eqnarray}
&&\chi_k \equiv  {1\over V} \rho_k ,
\nonumber \\ \
&& \chi_{k_1k_2} \equiv
{1\over V} \left( \rho_{k_1k_2} - \rho_{k_1}\rho_{k_2} \right) ,
\label{chidef}\\
&&\chi_{k_1k_2k_3} \equiv {1\over V} \left( 
\rho_{k_1k_2k_3} - \rho_{k_1k_2}\rho_{k_3} - \rho_{k_1k_3}\rho_{k_2} - \rho_{k_2k_3}\rho_{k_1}  
+ 2 \rho_{k_1}\rho_{k_2}\rho_{k_3} \right) ,
\nonumber
\end{eqnarray}
etc.... The $k$-point susceptibilities $\chi_k$ have already been introduced
in App.~\ref{notations}.
Analogously, beside the universal quantities $G_k$ defined in 
Eqs.~(\ref{Gkdef})
we consider the quantities
\begin{eqnarray}
&& G_{22} = \lim_{t\rightarrow 0^+}\;\lim_{V\rightarrow \infty}\;
\left[ - {\chi_{22} \over \xi^d \chi_2^2} \right],
\nonumber \\
&& G_{42} = \lim_{t\rightarrow 0^+}\;\lim_{V\rightarrow \infty}\;
\left[ - { \chi_{42} \over \xi^{2d} \chi_2^3}
+ 4 { \chi_4 \chi_{22} \over \xi^{2d} \chi_2^4}\right] ,
\nonumber\\
&&G_{222} = \lim_{t\rightarrow 0^+}\;\lim_{V\rightarrow \infty}\;
\left[ - { \chi_{222} \over \xi^{2d} \chi_2^3}
+ 6 { \chi_{22}^2 \over \xi^{2d} \chi_2^4 } \right] ,
\label{Gkkdef}
\end{eqnarray}
where $V$ is the volume.
In the FT approach one starts from the Hamiltonian
\begin{equation}
{\cal H}[\psi] = 
\int d^3x\, \left [ {1\over2} (\partial_\mu \phi)^2 + 
         {1\over2} (r + \psi(x)) \phi^2 + {v_0\over 4!} \phi^4 + 
         H \phi\right],
\end{equation}
where $\psi(x)$ is a spatially uncorrelated random field with Gaussian 
distribution. In the limit of small dilution this FT model should have 
the same critical behavior of the RIM \cite{replica}. Setting
\begin{equation}
\sigma^{\rm FT}_k = 
\left\langle \left[ \int d^3 x\, \phi(x) \right]^k\right\rangle_c,
\end{equation}
we define $\rho^{\rm FT}_{k_1,\ldots,k_n}$ and 
$\chi^{\rm FT}_{k_1,\ldots,k_n}$ by using Eqs.~(\ref{rhodef}) and 
(\ref{chidef}), with $\sigma_k$ replaced by $\sigma^{\rm FT}_k$.
Here the overline indicates the average over the 
random field $\psi(x)$. The universal constants $G_{k_1,\ldots,k_n}$
are given by the same expressions used in the spin model, 
Eqs.~(\ref{Gkdef}) and (\ref{Gkkdef}).

If $Z_\psi(H)$ is the partition function (the generator of the 
connected correlation functions in the FT language) for given
$H$ and disorder configuration $\psi$, we define
\begin{equation}
A_r(H_1,...,H_r) = 
  \overline{ {\ln} Z_\psi(H_1) \; ... {\ln} Z_\psi(H_r) },
\end{equation} 
so that
\begin{equation}
\rho_{k_1...k_r}^{\rm FT} = \left. {\partial^{k_1}\over \partial H_1^{k_1}}...
{\partial^{k_r}\over \partial H_r^{k_r}} A_r(H_1,...H_r)
\right|_{H_1=0,...H_r=0}.
\label{rhoFT}
\end{equation}
To compute these quantities we use the standard replica trick, i.e. we rewrite
\begin{equation}
{\rm ln} Z_\psi (H) = \lim_{N\rightarrow 0} {Z_\psi(H)^N-1\over N} .
\end{equation}  
Introducing $rN$ replicas, we write
\begin{equation}
A_r(H_1,...H_r) \approx
{1\over N^r} \int [d\psi] [d\varphi_{1,i}] \ldots [d\varphi_{r,i}]\, 
\exp \left[ - {\cal H}_{r\psi} + \sum_{k=1}^r H_k \sum_{i=1}^N \varphi_{k,i}
\right],
\label{arh}
\end{equation}
where
\begin{equation}
{\cal H}_{r\psi} =  \int d^3 x \left\{ {1\over 2} \sum_{ki}
      \left[ (\partial_\mu \varphi_{ki})^2 +  r \varphi_{ki}^2
+ \psi \varphi_{ki}^2 \right]
+{1\over 4!} \sum_{ki} v_0 \varphi_{ki}^4 \right\},
\label{H2psi}
\end{equation}
$k=1,...r$, and $i=1,..N$.
In Eq.~(\ref{arh}) we retained only the term depending
on all arguments since we only use this expression to
compute $\rho^{FT}_{k_1,\ldots,k_r}$ for $k_i > 0$ for all $i$.
Integrating out the disorder field $\psi$, we obtain
the generator of the connected correlation functions
\begin{equation}
A_r(H_1,...H_r) \propto
\int \prod [d\varphi_{k,i}]
\exp \left[ - {\cal H}_{r\varphi^4} +
\sum_{k=1}^r H_k \sum_{i=1}^N \varphi_{ki} \right],
\label{f2h1h2}
\end{equation}
where
\begin{equation}
{\cal H}_{r\varphi^4} =  \int d^3 x \left\{ {1\over 2} \sum_{ki}
      \left[ (\partial_\mu \varphi_{ki})^2 +  r \varphi_{ki}^2 \right]
+{1\over 4!} u_0 \left( \sum_{ki} \varphi_{ki}^2\right)^2
+{1\over 4!} v_0 \sum_{ki} \varphi_{ki}^4 \right\}\; .
\label{H2phi4}
\end{equation} 
The Hamiltonian (\ref{H2phi4}) is equivalent to the 
Hamiltonian (\ref{Hphi4rim}) with $N$ replaced by $rN$. 
In the limit $N\to 0$, this is of course irrelevant. Thus, if we define 
the $n$-point connected correlation function for the theory 
with Hamiltonian (\ref{Hphi4rim}), 
\begin{equation}
X_{a_1,\ldots,a_n}^{(n)} = 
  \left\langle \varphi_{a_1} \ldots \varphi_{a_n}\right \rangle_c,
\end{equation}
we obtain
\begin{eqnarray}
&&\chi_n = \lim_{N\rightarrow 0}  {1\over N} \sum_{a_1,\ldots,a_n=1}^N 
 X^{(n)}_{a_1\ldots a_n}, \\
&&\chi_{nm} = \lim_{N\rightarrow 0}
{1\over N^2} \sum_{a_1,\ldots,a_{n}=1}^N 
             \sum_{b_1,\ldots,b_{m}=N+1}^{2N} 
              X^{(n+m)}_{a_1\ldots a_{n}b_1\ldots b_{m}}, \\
&&\chi_{nmp} = \lim_{N\rightarrow 0}
{1\over N^3} 
     \sum_{a_1,\ldots,a_{n}=1}^N 
     \sum_{b_1,\ldots,b_{m}=N+1}^{2N} 
     \sum_{c_1,\ldots,c_{p}=2N+1}^{3N} 
           X^{(n+m+p)}_{a_1\ldots a_{n}b_1\ldots b_{m}c_1\ldots c_{p}} .
\end{eqnarray}
We wish finally to relate the constants $G_{k_1\ldots k_n}$ with the 
universal constants that parametrize the Helmoltz free energy of 
the FT model, cf. Eq.~(\ref{fec}). Using
\begin{eqnarray}
&& H_a = {\partial A_{\phi_4}\over \partial M_a}, \nonumber \\
&& X^{(n)}_{a_1 \ldots a_n} = 
   {\partial^{n-1} M_{a_1}\over \partial H_{a_2}\ldots \partial H_{a_n}},
\end{eqnarray}
we obtain 
\begin{eqnarray}
&&X^{(2)}_{ab} = \chi_2 \delta_{ab} ,\\
&&X^{(4)}_{abcd} = - \xi^d \chi_2^2
\left[ g_{41} {1\over 3} (\delta_{ab}\delta_{cd} + {\rm sym})
+ g_{42} \delta_{abcd}\right],
\nonumber \\
&&X^{(6)}_{abcdef} = - \xi^{2d} \chi_2^3
\Bigl[ 
(g_{61}-10 g_{41}^2) {1\over 15} 
(\delta_{ab}\delta_{cd}\delta_{ef} + {\rm sym}) + 
\nonumber \\
&&\qquad (g_{62}-20 g_{41} g_{42}) {1\over 15} 
( \delta_{abcd}\delta_{ef} + {\rm sym})+
(g_{63}-10 g_{42}^2) \delta_{abcdef} \Bigr],
\nonumber
\end{eqnarray}
where $\delta_{a_1...a_n}=1$ if all indices are equal and zero otherwise, 
and ``sym" indicates the appropriate permutations.

Then, using Eqs.~(\ref{Gkdef}) and (\ref{Gkkdef}), we find 
the following relations
\begin{eqnarray}
&&G_4 = g_{42} = v^*, \\
&&G_{22} = {1\over 3}g_{41}={1\over 3} u^*, \\
&&G_{6} = g_{63},\\
&&G_{42} = {1\over 15} g_{62}, \\
&&G_{222} = {1\over 15} g_{61}, \\
&&G_{8} = g_{85}.
\end{eqnarray}
Let us summarize the available numerical estimates for 
the above-reported  quantities.
The Monte Carlo simulations reported in Ref.~\cite{CMPV-03}
provided the results:
$G_4=43.3(2)$, $G_{22}=-6.1(1)$,
$r_6 = 0.90(15)$, $C_{42}\equiv G_{42}/(G_4 G_{22}) = 0.12(5)$, and 
$C_{222}\equiv G_{222}/G_{22}^2 = 0.45(15)$.
For the six-point couplings, we obtain correspondigly 
$G_6=1.7(3)\times 10^3$, $G_{42}=-32(13)$, $G_{222}=17(6)$.
The available FT estimates are more imprecise. 
For the four-point couplings, Ref.~\cite{PV-00} applied different resummation
methods to the six-loop $\beta$-functions, obtaining estimates that 
can be summarized by $G_4=38.0(1.5)$, $G_{22}=-4.5(6)$.
These estimates differ significantly from the Monte Carlo ones,
a discrepancy that is probably due to the non-Borel 
summability of the perturbative series \cite{nonBorel}.
Estimates of $G_6$ can be obtained from the results for 
$r_6$ reported in Sec.~\ref{FTappr}. By using the Monte 
Carlo estimate of $G_4$, we obtain $G_6 = 2100^{+200}_{-900}$.
Results less dependent on the location of the fixed point 
can be obtained by multiplying the estimate of $r_6$ at the 
FT (resp. Monte Carlo) fixed point for the corresponding FT (resp. 
Monte Carlo) estimate of $G_4$. At the FT fixed point $G_4 \approx 38$ 
and $r_6 = 1.1(1)$ so that $G_6 = 1.6(2)\times 10^3$,
while at the Monte Carlo fixed point $G_4 = 43.3(2)$ and
$r_6 = 0.6(3)$ that implies $G_6 = 1.1(5)\times 10^3$. Taking as final
estimate that derived by using the FT estimate of $G_4$, we have 
$G_6 = 1600^{+200}_{-500}$, where the error is such to include also 
the estimate at the Monte Carlo fixed point. We also used field theory 
to determine the replica-replica six-point couplings. 
We applied the Pad\'e-Borel method to the four-loop perturbative 
expansions \cite{PS-01,PS-02} of $C_{42} \equiv G_{42}/(G_{22} G_4)$ and of 
$C_{222} \equiv G_{222}/G_{22}^2$. 
The results are only indicative since the estimates change significantly
with the order and with the Pad\'e approximant. 
We obtain $C_{42} = 0.06(3)$, 0.01(5) and $C_{222} = 0.05(2)$, $-0.015(25)$
by using the FT and the Monte Carlo estimate of the fixed point 
respectively. Comparing with the Monte Carlo results reported above
we observe that field theory (at four loops)
provides only the order of magnitude, but is 
unable to be quantitatively predictive. Finally, we consider $r_8$ and $G_8$.
As discussed in Sec.~\ref{FTappr} (see also Ref. \cite{PBanr8}), 
the perturbative three-loop expansion
of $r_8$ gives only an upper bound. 
A much more precise estimate of $r_8$ has been derived in Sec.~\ref{reseq},
$r_8 = 1.5(5)$, which implies $G_8=1.2(4)\times 10^5$.

\section{Field-theoretical determination of $R_\xi^+$}
\label{specheat}

In this appendix we estimate the universal amplitude ratio
$R_\xi^+= f^+ (\alpha A^+)^{1/3}$ 
by performing a six-loop expansion in the framework of a fixed-dimension FT 
approach based on the Hamiltonian (\ref{Hphi4rim}).
In this scheme the theory is renormalized by introducing
a set of zero-momentum conditions for the one-particle irreducible 
two-point and four-point correlation functions:
\begin{eqnarray}
&&\Gamma^{(2)}_{ab}(p) =
  \delta_{ab} Z_\varphi^{-1} \left[ m^2+p^2+O(p^4)\right],
\label{ren1}  \\
&&\Gamma^{(4)}_{abcd}(0) = 
Z_\varphi^{-2} m \left( u S_{abcd} + v C_{abcd} \right),
\label{ren2}  
\end{eqnarray}
where 
$S_{abcd}=\case{1}{3} \left(\delta_{ab}\delta_{cd} + \delta_{ac}\delta_{bd} + 
                \delta_{ad}\delta_{bc} \right)$
and $C_{abcd}=\delta_{ab}\delta_{ac}\delta_{ad}$.
Eqs.~(\ref{ren1}) and (\ref{ren2})
relate the mass scale $m$ and the zero-momentum
quartic couplings $u$ and $v$ to the corresponding Hamiltonian parameters
$r$, $u_0$, and $v_0$,
\begin{equation}
u_0 = m u Z_u Z_\varphi^{-2},\qquad v_0 = m v Z_v Z_\varphi^{-2}.
\label{u0v0}
\end{equation}
In addition one defines the function $Z_t$ through the relation
\begin{equation}
\Gamma^{(1,2)}_{ab}(0) = \delta_{ab} Z_t^{-1},
\label{zt}
\end{equation}
where $\Gamma^{(1,2)}$ is the one-particle irreducible
two-point function with an insertion of $\case{1}{2}\phi^2$.
The pertubative expansions of $Z_\varphi(u,v)$, 
$Z_u(u,v)$, $Z_v(u,v)$, $Z_t(u,v)$
have been computed to six loops \cite{CPV-00,PV-00}.
The specific heat is given by the zero-momentum energy-energy
correlation function averaged over the random dilution.
In the FT approach it corresponds to
\begin{equation}
C_H = \lim_{N\rightarrow 0} {1\over N}
\int d^d x \; \langle \varphi^2(0) \varphi^2(x) \rangle_c,
\label{CCCC}
\end{equation}
i.e. to the zero-momentum value of the two-point
correlation function of the operator $\varphi^2=\sum_a \varphi_a^2$.
We computed $C_H$ to six loops, extending the 
five-loop computation of Ref.~\cite{Mayer-98}.
The calculation requires
the evaluation of a few hundred Feynman  diagrams.
We handled it with a symbolic manipulation program, which  generates the diagrams 
and computes the symmetry and group factors of each of them.
We used the numerical results compiled in Ref.~\cite{NMB-77}
for the integrals associated with each diagram.

In order to compute the universal ratio $R_\xi^+$, we follow Ref.~\cite{BS-92}.
We consider different expansions that converge to $R_\xi^+$ as $t\to 0^+$:
\begin{eqnarray}
&&R^{(1)}_\xi = \lim_{t\rightarrow 0^+}
\left[ \gamma \xi ( -C_H')^{1/d} 
\left( {d {\rm ln} \chi\over dt} \right)^{-1} \right] ,\\
&&R^{(2)}_\xi = \lim_{t\rightarrow 0^+}
\left\{ {d\over dt} 
\left[ \xi^{-1} ( - C_H')^{-1/d} \right] \right\}^{-1} ,
\end{eqnarray}
where $C_H'\equiv d C_H/dt$ (at fixed $u_0$ and $v_0$) 
and $t$ is the reduced temperature.
The derivatives with respect to the reduced temperature
$t$ can be done by using Eq.~(\ref{zt}), which can be rewritten as
\begin{equation}
\left. {d \chi^{-1}\over dt} \right|_{u_0,v_0}= Z_t^{-1}.
\end{equation}
This allows us to compute the derivative with respect to 
$t$ of generic functions written in terms of $u_0, v_0$, and $\chi$.
For example, setting 
\begin{equation}
C_H = \chi^{1/2} \overline{C}_H(u_1,v_1),
\qquad
u_1=\chi^{1/2}u_0 ,\quad v_1=\chi^{1/2}v_0,
\end{equation}
we obtain
\begin{equation}
C_H'(t) = - {1\over 2} \chi^{3/2} Z_t^{-1} 
\left( 1 + u_1 {\partial\over \partial u_1 } + 
v_1 {\partial\over \partial v_1} \right) \overline{C}_H(u_1,v_1).
\end{equation}
Using the relations $\xi=1/m$,
$\chi=Z_\varphi \xi^2$, and Eq.~(\ref{u0v0}), one obtains expressions
that can be expanded in powers of the renormalized quartic couplings
$u$ and $v$.
For example,
\begin{equation}
R^{(1)}_\xi(u,v) = {1\over 2} \gamma Z_\varphi^{-1/2} Z_t^{2/3}
\left( 1 + u_1 {\partial\over \partial u_1 } + 
v_1 {\partial\over \partial v_1} \right) 
\overline{C}_H(u_1,v_1),
\end{equation}
where $u_1=Z_u Z_\varphi^{-1} u$, $v_1=Z_v Z_\varphi^{-1} v$.
We write
\begin{equation}
R^{(n)}_\xi(u,v) = \sum_{i,j} c_{ij}^{(n)} \bar{u}^i \bar{v}^j,
\end{equation}
where $\bar{u}$ and $\bar{v}$
are the rescaled couplings
$\bar{u}\equiv u/(6\pi)$ and $\bar{v}\equiv  3 v/(16\pi)$.
The coefficients $c_{ij}^{(n)}$ are reported
in Table~\ref{rxico} for $0\leq i+j\leq  5$.
Estimates of $R_\xi^+$ are obtained
by resumming these series, and evaluating them at $u^*$ and $v^*$.
We employed several resummation methods, see,
e.g., Refs.~\cite{PV-00,CMPV-03}.
Without entering into the details, we report the results:
$R^{(1)}_\xi = 0.287(2)$ and $R^{(2)}_\xi=0.286(3)$
obtained by using the FT estimates of the fixed point,
$u^* = 37$ and $v^* = -13$, 
and  $R^{(1)}_\xi = 0.296(2)$ and $R^{(2)}_\xi=0.295(2)$
obtained by
using the Monte Carlo estimates \cite{CMPV-03}
$u^*=-18.6(3)$ and $v^*=43.3(2)$.
Our final estimate is $R_\xi^+=0.290(10)$
that includes all the above-reported results.
Our FT estimate agrees  
with the Monte Carlo result of Ref.~\cite{CMPV-03}, $R_\xi^+=0.2885(15)$,
and with the FT estimates of Refs.~\cite{BS-92,Mayer-98},
$R_\xi^+=0.286(4)$ (four loops) and
$R_\xi^+\approx 0.2887$ (five loops).

\begin{table}[tbp]
\caption{
Coefficients $c^{(1)}_{ij}$ and $c^{(2)}_{ij}$ of the expansion of
$R^{(1)}_\xi$ and $R^{(2)}_\xi$ respectively. 
\label{rxico}
}
\begin{tabular}{cll}
\multicolumn{1}{c}{$i,j$}&
\multicolumn{1}{c}{$c^{(1)}_{ij}$}&
\multicolumn{1}{c}{$c^{(2)}_{ij}$}\\
\tableline
0,0  &   0.2150635  &     0.2150635 \\
0,1  &   0.0358439  &     0.0358439 \\
1,0  &   0.0268829  &     0.0268829 \\
0,2  &   0.0000532  &  $-$0.0019913 \\
1,1  &$-$0.0043608  &  $-$0.0089610 \\
2,0  &$-$0.0016353  &  $-$0.0033604 \\
0,3  &   0.0025329  &     0.0055607 \\
1,2  &   0.0079130  &     0.0183873 \\
2,1  &   0.0079586  &     0.0187523 \\
3,0  &   0.0019897  &     0.0046881 \\
0,4  &$-$0.0021046  &  $-$0.0052579 \\
1,3  &$-$0.0097350  &  $-$0.0242605 \\
2,2  &$-$0.0167427  &  $-$0.0414335 \\
3,1  &$-$0.0110331  &  $-$0.0273141 \\
4,0  &$-$0.0020687  &  $-$0.0051214 \\
0,5  &   0.0024689  &     0.0072264 \\
1,4  &   0.0140199  &     0.0411159 \\
2,3  &   0.0316913  &     0.0929392 \\
3,2  &   0.0345501  &     0.1009750 \\
4,1  &   0.0165332  &     0.0482474 \\
5,0  &   0.0024800  &     0.0072371 \\
\end{tabular}
\end{table}



\begin{references}
\bibitem{PBanr8}
The estimate $r_8=15(15)$ has been obtained 
by means of a standard Pad\'e-Borel analysis (more sophisticated analyses, 
such as those described in Ref.~\cite{PV-00}, cannot be used here because the 
perturbative series is too short):
[2/1] approximants give $r_8\approx 10$ and $r_8\approx 26$ respectively
at the FT and Monte Carlo estimates of the
fixed point, while [1/1] ones give
$r_8\approx -1$ and $r_8\approx 1$ respectively.
\end{references}
\end{document}